\documentclass[12pt]{article}
\usepackage{amsfonts}

\newcommand{\Z}{{\Bbb Z}}

\newcommand{\io}{[\hspace{-1pt}[}
\newcommand{\ic}{]\hspace{-1pt}]}
\newcommand{\fo}{\{\!\mid\!}
\newcommand{\fc}{\!\mid\!\}}

\newcommand{\sgn}{{\rm sgn}}

\newcommand{\ren}{{\rm ren}}
\newcommand{\tr}{{\rm tr}}

\renewcommand{\arctan}{{\rm arctan}}

\renewcommand{\cosh}{{\rm cosh}}

\def\be{\begin{equation}}
\def\ee{\end{equation}}
\def\bea{\begin{eqnarray}}
\def\eea{\end{eqnarray}}
\def\ba{\begin{array}}
\def\ea{\end{array}}
\def\beq{\begin{eqnarray*}}
\def\eeq{\end{eqnarray*}}

\def\I{{\rm I}}

\def\vac{{\rm vac}}
\def\x{{\bf x}}

\def\C{{\cal C}}

\begin{document}

\title{Chiral symmetry breaking as a consequence of nontrivial spatial topology}
\author{\bf{Yurii A. Sitenko}\thanks{Electronic addresses: yusitenko@bitp.kiev.ua; 
sitenko@itp.unibe.ch}\\
Bogolyubov Institute for Theoretical Physics,\\
National Academy of Sciences,\\
14-b Metrologichna str., 252143, Kyiv, Ukraine\thanks{Permanent address}\\
and\\
Institute for Theoretical Physics\\
Berne University\\
Sidlerstrasse 5, 3012 Berne, Switzerland}
\date{}
\maketitle

\begin{abstract}
A singular configuration of external static magnetic field in the form
of a pointlike vortex polarizes the vacuum of quantized massless
spinor field in 2+1-dimensional space-time. This results in an analogue
of the Bohm-Aharonov effect: the chiral symmetry breaking
condensate, energy density and current emerge in the vacuum even in the
case when the spatial region of nonvanishing external field strength is
excluded. The dependence of the vacuum characteristics both on the
value of the vortex flux and on the choice of the boundary condition at
the location of the vortex is determined.\\
Keywords: vacuum condensate, chiral symmetry. singular vortex
\end{abstract}

The dynamical breakdown of chiral symmetry is a topic of
persisting interest for a long time (for reviews see, e.g., 
Refs.\cite{Hig,Mir,Hats,Inag}). An important role in its study is played by
the appropriate "order parameter" -- vacuum condensate

\be
\C(\x,t)=i\langle \vac|T\bar{\Psi}(\x,t)\Psi(\x,t)|\vac\rangle.
\ee
In the background of external classical fields, the vacuum
expectation value of the time-ordered product of the fermion
field operators takes the form

\be
\langle\vac| T\Psi(\x,t)\bar{\Psi}(\x',t')|\vac\rangle = -i\langle\x,t|
(-i\gamma^\mu\nabla_\mu+m)^{-1} |\x',t'\rangle,
\ee
where $\nabla_\mu$ is the covariant derivative in this background.
Thus, in a static background, condensate (1) is reduced to the form
\be
\C(\x)=-{1\over2} \tr\langle\x|\gamma^0\,\sgn(H)|\x\rangle,
\ee
where
\be
H=-i\gamma^0\gamma^j\nabla_j+\gamma^0m,
\ee
is the pertinent Dirac Hamiltonian and
\beq
\sgn(u)=\left\{\ba{cc}
1,& u>0\\ -1,&u<0\\ \ea \right\}.
\eeq
For a certain background field (magnetic or gravitational)
configuration, the analysis of Eq.(1) (or its particular form,
Eq.(3)) can be carried out with a special attention to the
limiting procedure $m\rightarrow 0$. If the condensate survives
in this limit, then it witnesses chiral symmetry breaking by this
configuration. Moreover, as it has been shown in Ref.\cite{Gus}, 
in the case of quantized spinor fields in the
background of a homogeneous magnetic field in 2+1-dimensional
space-time, the chiral symmetry breaking condensate emerges
irrespectively of all other possible types of interaction among
quantized spinor fields. The question that we would like to
address in the present Letter is, whether the emergence of the
chiral symmetry breaking condensate could be caused by nontrivial
topology of the base space manifold?

Nontrivial topology can be achieved, formally, by excluding
submanifolds of less dimensions from an initial manifold with
trivial topology. For example, deleting a line from a
three-dimensional space results in the spatial topology becoming
nontrivial with the classification in terms of the winding
number: $\pi_1=\Z$ (here $\pi_1$ is the first homotopy group and
$\Z$ is the set of integer numbers). Imposing various boundary
conditions at the location of the excluded submanifold, one can
study a possibility that some of them will induce the chiral
symmetry breaking condensate in the vacuum. However, our freedom
to vary boundary conditions is restricted by the natural
requirement of their physical meaningfulness. Since we are not to
consider  here boundary conditions invoking instability and
confine ourselves to stationary problems, self-adjointness of the
Hamiltonian has to be maintained. It is well known that the free
Dirac Hamiltonian is {\it essentially} self-adjoint when defined on the
domain of regular functions (see, e.g., Ref.\cite{Dir1}). Thus,
we have no choice but to pick the regularity boundary condition
in order to ensure self-adjointness, and this yields no gain
comparing to the case of trivial topology.  One has to do
something more, than simply saying about the deletion of a line
from a space, in order to achieve nontrivial topology physically.
And this doing more means inserting a singular magnetic vortex at
the location of the deleted line. Then the Dirac Hamiltonian
defined on the domain of regular functions is not self-adjoint
any more, which may seem from the first sight to be rather
forbidding.  However, this obstruction can be overcome by
exploring a possibility of self-adjoint {\it extension} along the lines
of the Weyl-von Neumann theory of self-adjoint operators (see,
e.g., Refs.\cite{Akhie,Reed}). As a result, one arrives at a
set of boundary conditions which are compatible with
self-adjointness and allow for  wave
functions  being square-integrable but irregular at the location
of a deleted line. We shall show that these boundary
conditions can induce the chiral symmetry breaking condensate in
the vacuum.

Omitting the spatial dimension along a deleted line, one gets a
two-dimensional space (plane) with a deleted point (puncture). A
static singular magnetic vortex inserted at the puncture is given
by the Ehrenberg-Siday-Aharonov-Bohm potential \cite{Ehre,Aha}

\be
V^1=-\Phi {x^2\over (x^1)^2+(x^2)^2}, \quad V^2=\Phi{x^1\over
(x^1)^2+(x^2)^2},
\ee
where $\Phi$ is the vortex flux in $2\pi$  units (conventions
$\hbar=c=1$ are implied) and the location of the puncture is
taken as the origin of the Carthesian coordinates $x^1$ and $x^2$
on the plane. Specifying the covariant derivative as

\be
\nabla_j=\partial_j+i V_j, \quad j=1,2,
\ee
our aim is to study the polarization of the fermionic vacuum by
classical static background (5) in 2+1-dimensional space-time.

As it is known (see, e.g., Ref.\cite{Appe}), in order to
consider the chiral symmetry breaking in 2+1 dimensions,
quantized spinor fields have to be assigned to the reducible
representation of the Clifford algebra
\be
\Psi=\left( \ba{c} \Psi_+\\ \Psi_-\\ \ea \right),
\ee
where $\Psi_+$ and $\Psi_-$ are assigned to two inequivalent
irreducible ($2\times2$) representations; $\gamma$ matrices are chosen
in the form

\be
\gamma^0=\left(\ba{cc}
\sigma_3& 0\\ 0& -\sigma_3\\ \ea \right),
\quad
\gamma^1=\left(\ba{cc}
i\sigma_1& 0\\ 0& -i\sigma_1\\ \ea \right),
\quad
\gamma^2=\left(\ba{cc}
i\sigma_2& 0\\ 0& -i\sigma_2\\ \ea \right),
\ee
where $\sigma_1,\sigma_2$ and $\sigma_3$ are the Pauli matrices. The
algebra is completed by adding the $\gamma$ matrix corresponding to the
missing dimension,

\be
\gamma^3=i\left(\ba{cc}
 0&\I\\ \I& 0\\ \ea \right),
\ee
and the $\gamma^5$ matrix is
\be
\gamma^5 \equiv i\gamma^0\gamma^1\gamma^2\gamma^3
=i\left(\ba{cc}
 0&\I\\
-\I& 0\\ \ea \right).
\ee
Then the massless Dirac Hamiltonian (Eq.(4) at $m=0$) in background (5)
takes the form

\be
H=\left(\ba{cc}
H_+& 0\\
0& H_-\\ \ea \right),
\ee
where
\be
H_\pm=\left(\ba{cc}
0& e^{-i\varphi}[\partial_r-r^{-1}(i\partial_\varphi+\Phi)]\\
e^{i\varphi}[-\partial_r-r^{-1}(i\partial_\varphi+\Phi)]&0\\ \ea
\right),
\ee
$r=\sqrt{(x^1)^2+(x^2)^2}$ and $\varphi=\arctan(x^2/x^1)$ are the polar
coordinates. A single-valued solution to the stationary Dirac equation
\be
H\langle \x|E\rangle =E\langle\x|E\rangle
\ee
with Hamiltonian (11)--(12) is presented as

\be
\langle\x|E\rangle=\sum_{n\in\Z}
\left(\ba{c}
f_n^+(r,E)e^{in\varphi}\\
g_n^+(r,E)e^{i(n+1)\varphi}\\
f_n^-(r,E)e^{in\varphi}\\
g_n^-(r,E)e^{i(n+1)\varphi}\\ \ea
\right).
\ee

\renewcommand{\footnote}[1]{%
\def\thefootnote{\arabic{footnote})}%
\footnotemark\footnotetext{#1}}

Decomposing the value of the vortex flux into the integer and
fractional parts,
\be
\Phi=\io\Phi\ic+\fo \Phi\fc, \quad 0\leq\fo\Phi\fc<1
\ee
($\io u\ic$ denotes the integer part of quantity $u$), one can note
that the case of $\fo \Phi\fc=0$ is equivalent to the case of trivial
topology, i.e. absence of the vortex $(\Phi=0)$.\footnote{This confirms
once more the general fact that a singular magnetic vortex is
physically unobservbale at integer values of the vortex flux
\cite{Aha}. It was as far back as 1931 that Dirac used actually this
fact to obtain his famous condition for the magnetic monopole
quantization \cite{Dir2}.} In the case of $0<\fo \Phi\fc<1$ the modes
with $n\neq\io\Phi\ic$ obey the condition of regularity at the puncture
$r=0$: the corresponding partial Hamiltonians are essentially
self-adjoint when defined on the domain of regular functions. That is
only the partial Hamiltonian corresponding to $n=\io \Phi\ic$, that
needs a self-adjoint extension. Similarly to the case of nonzero
fermion mass ($m\neq0$) \cite{Ger,Sit96,Sit97}, it can be shown that
the self-adjoint extension is parametrized by one real continuous
variable $\Theta$, yielding the following condition for the mode with
$n=\io \Phi\ic$:

\be
\cos\left({\Theta\over2}+{\pi\over4}\right)\lim_{r\rightarrow 0}\,
(\mu r)^{\fo \,\Phi\,\fc}f^\pm_{\io \Phi\ic}=\mp
\sin\left({\Theta\over2}+{\pi\over4}\right)
\lim_{r\rightarrow 0}\,(\mu r)^{1-\fo \,\Phi\,\fc}g^\pm_{\io \Phi\ic},
\ee
where $\mu>0$ is the parameter of the dimension of inverse length,
which is introduced just to scale the different irregular behaviour of
the $f$ and $g$ components; note that, since Eq.(16) is periodic in
$\Theta$ with period $2\pi$, all permissible values of $\Theta$ can be
restricted, without a loss of generality, to range
$-\pi\leq\Theta\leq\Theta$.

The "minus/plus" sign factor in the right-hand side of Eq.(16) deserves
a special comment. For a massless fermion in an irreducible
representation, an overall change of sign before the Pauli matrices
does not mean a transition to an inequivalent representation, because
$\sigma_3$ by itself does not appear anywhere, while a change of sign
before $\sigma_1$ and $\sigma_2$ corresponds to a transition to an
equivalent representation. This is reflected by the fact that the
expressions for $H_+$ and $H_-$ coincide, see Eq.(12). Thus, boundary
condition (16) is the only point where a distinction between $\Psi_+$
and $\Psi_-$ arises. Unless we had made this distinction, we would get
the case of a reducible representation composed of two equivalent
irreducible ones, which simply doubles the case of an irreducible
representation considered in detail elsewhere \cite{Sit99}: the
condensate and the corresponding mass term break parity rather than
chiral symmetry in that case. On the contrary, boundary condition (16)
is invariant under parity transformation

\be
\langle x^1,x^2|E\rangle \rightarrow i\gamma^1\gamma^3\langle
-x^1,x^2|E\rangle
\ee
and is not invariant under chiral transformation
\be
\langle x^1,x^2|E\rangle \rightarrow \gamma^5\langle
x^1,x^2|E\rangle,
\ee
unless
\be
\cos\Theta=0,
\ee
when both parity and chiral symmetry are maintained.

Having specified the boundary condition at the puncture $r=0$, an
explicit form of a solution to Eq.(13) can be obtained. Using this
form, all vacuum polarization effects are determined. In addition to
condensate (3), also current

\be
j_\varphi(\x)=-{1\over2}\tr
\langle\x|\gamma^0(\gamma^2\cos\varphi-\gamma^1\sin\varphi)
\,\sgn(H)|\x\rangle
\ee
and energy density

\be
\varepsilon^\ren(\x)=-{1\over2}\tr\langle \x|\,|H|\,|\x\rangle^{(\ren)}
\ee
are induced in the vacuum. Here, the superscript "ren" in Eq.(21)
denotes the use of a certain regularization and renormalization
procedure to get rid of both ultraviolet and infrared divergences, for
details see Ref.\cite{Sit99}. Concerning Eq.(20), we can add that
the radial component of the vacuum current,

\be
j_r(\x)=-{1\over2}\tr
\langle\x|\gamma^0(\gamma^1\cos\varphi+\gamma^2\sin\varphi)\,\sgn
(H)|\x\rangle,
\ee
is not induced.

We list below the results (details will be published elsewhere):
\be
\C(\x)=-{\sin(\fo \Phi\fc\pi)\over \pi^3r^2}\int\limits_0^\infty dw\,
w{K^2_{\fo \,\Phi\,\fc}(w)+K^2_{1-\fo \,\Phi\,\fc}(w)
\over \cosh[(2\fo \Phi\fc-1)\ln \bigl({w\over\mu
r}\bigr)+\ln\bar{A}]},
\ee

\[
j_\varphi(\x)  =  {\sin(\fo \Phi\fc\pi)\over \pi r^2}
\left\{ {(\fo \Phi\fc-{1\over2})^2\over 2\cos(\fo \Phi\fc\pi)} 
\right.-
\]
\be
-\left.{2\over \pi^2}\int\limits_0^\infty dw\, w
K_{\fo \,\Phi\,\fc}(w)K_{1-\fo \,\Phi\,\fc}(w) \tanh\biggl[
(2\fo \Phi\fc-1)\ln\bigl({w\over\mu
r}\bigr)+\ln\bar{A}\biggr]\right\}, 
\ee

\[
\varepsilon^\ren(\x)  =  {\sin(\fo \Phi\fc\pi)\over \pi
r^3} \left\{ {{1\over2}-\fo \Phi\fc\over 6\cos(\fo \Phi\fc\pi)}
\left[{3\over4}-\fo \Phi\fc(1-\fo \Phi\fc)\right]+\right.\]
\be + \left. {1\over\pi^2} \int\limits_0^\infty dw\,
w^2\bigl[K^2_{\fo \,\Phi\,\fc}(w) -K^2_{1-\fo \,\Phi\,\fc}(w) \bigr]
\tanh\biggl[(2\fo \Phi\fc-1)\ln\bigl({w\over\mu
r}\bigr)+\ln\bar{A}\biggr]\right\},
\ee
where
\be
\bar{A}=2^{1-2\fo \,\Phi\,\fc}\, {\Gamma(1-\fo \Phi\fc)
\over\Gamma(\fo \Phi\fc)} \tan\bigl({\Theta\over2}+{\pi\over4}\bigr),
\ee
$K_\tau(w)$ is the Macdonald function of order $\tau$, and $\Gamma(u)$
is the Euler gamma function.

Since the puncture breaks translational invariance on the plane, the
vacuum polarization effects (23) -- (25) are not translationally
invariant. There remains an invariance with respect to rotations around
the puncture, and, therefore, the vacuum polarization effects depend
only on the distance from the puncture. At large distances they are
decreasing by power law:

\be
\C(\x)
\mathop{=}\limits_{r\rightarrow \infty}
-{\sin(\fo \Phi\fc\pi)\over \pi^2r^2} \left\{
\ba{cc}
(\mu r)^{2\fo \,\Phi\,\fc-1} \bar{A}^{-1} {\Gamma\bigl({3\over2}-
\fo \,\Phi\,\fc\bigr)\Gamma\bigl({3\over2}-2\fo \,\Phi\,\fc\bigr)\over
\Gamma(1-\fo \,\Phi\,\fc)}, & 0<\fo \Phi\fc<{1\over2}\\[0.4cm]
(\mu r)^{1-2\fo \,\Phi\,\fc}\bar{A} {\Gamma(\fo \,\Phi\,\fc+{1\over2})
\Gamma(2\fo \,\Phi\,\fc-{1\over2})\over \Gamma(\fo \,\Phi\,\fc)},&
{1\over2}<\fo \Phi\fc<1\\ \ea \right. \, ,
\ee

\be
j_\varphi(\x)
\mathop{=}\limits_{r\rightarrow \infty}
{\tan(\fo \Phi\fc\pi)\over2\pi r^2} |\fo \Phi\fc-{1\over2}| \left(|
\fo \Phi\fc-{1\over2}|-1\right),
\ee

\be
\varepsilon^\ren(\x)
\mathop{=}\limits_{r\rightarrow \infty}
{\tan(\fo \Phi\fc\pi)\over2\pi r^3} \biggl(\fo \Phi\fc-{1\over2} \biggr)
\left[{1\over3}\fo \Phi\fc(1-\fo \Phi\fc)-{1\over4}+{1\over2}|
\fo \Phi\fc-{1\over2}|\right].
\ee
At half-integer values of the vortex flux, taking into account relation

\be
\bar{A}|_{\fo \,\Phi\,\fc ={1\over2}}=
\tan\bigl({\Theta\over2}+{\pi\over4}\bigr),
\ee
we get
\be
\C(\x)|_{\fo \,\Phi\,\fc ={1\over2}}=-{\cos\Theta\over 2\pi^2 r^2},
\ee

\be
j_\varphi(\x)|_{\fo \,\Phi\,\fc ={1\over2}}=-{\sin\Theta\over 2\pi^2r^2},
\ee

\be
\varepsilon^\ren(\x)|_{\fo \,\Phi\,\fc ={1\over2}}={1\over12\pi^2r^3};
\ee
note the $\Theta$ independence of the last relation.

Integrating Eq.(23) over the whole plane, we obtain the total vacuum
condensate

\be
\C\equiv \int d^2x\, \C(\x)=-{\sgn_0(\cos\Theta)\over
|2\fo\Phi\fc -1|}
\ee
where
\beq
\sgn_0(u)=\left\{\ba{cc}
\sgn(u),& u\neq0\\
0,& u=0\\ \ea \right\}.
\eeq
Thus, the total vacuum condensate is infinite at half-integer values of
the vortex flux, unless Eq.(19) holds.

Finally, few comments on the case of absence of the condensate in the
vacuum, i.e., when Eq.(19) holds. In this case two of the four
components of the wave function (14) become regular for all $n$: if
$\Theta={\pi\over2}$, then the $g_n^\pm$ components are regular, and,
if $\Theta=-{\pi\over2}$, then the $f_n^\pm$ components are regular.
Due to this, scale symmetry, as well as parity and chiral symmetry, is
maintained. The nontrivial topology reveals itself in the emergence of
the vacuum current and energy density only:

\bea
j_\varphi(\x) & = & {\tan(\fo\Phi\fc \pi)\over 2\pi r^2}
\bigl( \fo\Phi\fc -{1\over2}\bigr) \bigl(\fo\Phi\fc
-{1\over2}\pm1\bigr), \quad \Theta=\pm{\pi\over2}, \\[0.3cm]
\varepsilon^\ren(\x) & = & {\tan(\fo\Phi\fc \pi) \over2\pi
r^3} \bigl(\fo\Phi\fc -{1\over2}\bigr) \left[{1\over3}\fo\Phi\fc
\bigl(1-\fo\Phi\fc \bigr)-{1\over4}\mp {1\over2}\bigl(\fo\Phi\fc
-{1\over2}\bigr) \right], \nonumber\\[0.3cm]
&&\Theta=\pm{\pi\over2}.
\eea

Summarizing, we have completed an exhaustive study of vacuum
polarization effects in the background  of a singular magnetic vortex
$(\fo\Phi\fc\neq0)$ under boundary condition (16) which ensures
self-adjointness of the Hamiltonian and parity conservation. If
$\Theta\neq\pm{\pi\over2}$, then chiral symmetry breaking condensate
(23) emerges in the vacuum. One would anticipate that scale symmetry is
broken as well, but this is not the case. Note that  the vacuum current
and energy density contain scale invariant pieces (terms which are not
represented by integrals in Eqs.(24) and (25)).\footnote{In
2+1-dimensional space-time the canonical dimension of the fermion
field operator is equal to one in the units of the power of inverse
length.}
Also, large distance asymptotics of the vacuum current and
energy density are scale invariant, see Eqs.(28) and (29). But the most
significant point is that, at half-integer values of the vortex flux
$(\fo\Phi\fc={1\over2}$), the vacuum condensate, as well as the vacuum
current and energy density, becomes scale invariant, see Eqs.(31) --
(33); consequently, the total vacuum condensate becomes infinite, see
Eq.(34). That is only at non-half-integer values of the vortex flux,
that both chiral and scale symmetries are broken.

In conclusion, it should be emphasized once more that magnetic field
strength  $\partial_1V_2-\partial_2V_1$ corresponding to potential (5)
vanishes everywhere on the plane punctured at $x^1=x^2=0$. Thus, we
have shown that, in the absence of any background field strength or
curvature, chiral symmetry breaking occurs just due to nontrivial
topology of space (in the Bohm-Aharonov-effect-like manner). It would
be interesting to bring possible types of interaction among quantized
fermion fields (either four-fermionic, or electrodynamic) into this
context in order to consider the mass gap equation and its
implications.


\section*{Acknowledgements}

I am grateful to H.~Leutwyler and V.A.~Miransky for stimulating
discussions and interesting remarks. The research was supported by the
State Foundation for Fundamental Research of Ukraine (project
2.4/320) and the Swiss National Science Foundation (grant
CEEC/NIS/96-98/7 IP 051219).


\begin{thebibliography}{99}
\bibitem{Hig} K.~Higashijima, Progr.Theor.Phys.Suppl. {\bf 104}, 1
(1991).
\bibitem{Mir} V.A.~Miransky, Dynamical Symmetry Breaking in Quantum
Field Theories. (World Scientific, Singapore, 1993).
\bibitem{Hats} T.~Hatsuda and T.~Kunihiro, Phys.Rep. {\bf 247},
221 (1994).
\bibitem{Inag} T.~Inagaki, T.~Muta and S.D.~Odintsov,
Progr. Theor. Phys. Suppl. {\bf 127}, 93 (1997).
\bibitem{Gus} V.P.~Gusynin, V.A.~Miransky and I.A.~Shovkovy, Phys.
Rev. Lett. {\bf 73}, 3499 (1994); Phys. Rev. {\bf D52}, 4718 (1995).
\bibitem{Dir1} B.~Thaller, The Dirac Equation (Springer-Verlag, Berlin,
1992).
\bibitem{Akhie} N.I.~Akhiezer and I.M.~Glazman, Theory of Linear
Operators in Hilbert Space (Pitman, Boston, 1981) V.2.
\bibitem{Reed} M.~Reed and B.~Simon, Methods of Modern Mathematical
Physics, V.2: Fourier Analysis, Self-Adjointness (Academic Press, New
York, 1975).
\bibitem{Ehre} W.~Ehrenberg and R.Siday, Proc. Phys. Soc. (London) {\bf
B62}, 8 (1949).
\bibitem{Aha} Y.~Aharonov and D.~Bohm, Phys. Rev. {\bf 115}, 485
(1959).
\bibitem{Appe} T.W.~Appelquist, M.~Bowick, D.~Karabali and
L.C.R.~Wijewardhara, Phys. Rev. {\bf D33}, 3704 (1986).
\bibitem{Dir2} P.A.M.~Dirac, Proc. Roy. Soc. (London) {\bf A133}, 60
(1931).
\bibitem{Ger} Ph. de S.~Gerbert, Phys. Rev. {\bf D40}, 1346 (1989).
\bibitem{Sit96} Yu.A.~Sitenko, Phys. Lett. {\bf B387}, 334 (1996).
\bibitem{Sit97} Yu.A.~Sitenko, Phys. Atom. Nucl. {\bf 60}, 2102
(1997).
\bibitem{Sit99} Yu.A.~Sitenko, to be published.
\end{thebibliography}
\end{document}